\documentclass{aa}
\usepackage{txfonts}
\usepackage{graphicx}
\usepackage[latin1]{inputenc}

\def\HII{H\,{\sc{ii}}}

\def\fs{\hbox{$.\!\!^{\rm s}$}}

\def\farcm{\hbox{$.\mkern-4mu^\prime$}}

\def\arcmin{\hbox{$^\prime$}}
\def\arcsec{\hbox{$^{\prime\prime}$}}

\def\degr{\hbox{$^\circ$}}
\def\h{\hbox{$^{\reset@font\r@mn{h}}$}}
\def\m{\hbox{$^{\reset@font\r@mn{m}}$}}
\def\s{\hbox{$^{\reset@font\r@mn{s}}$}}
\def\msol{\hbox{\kern 0.20em $M_\odot$}}

\def\kms{\hbox{\kern 0.20em km\kern 0.20em s$^{-1}$}}
\def\cmmt{\hbox{\kern 0.20em cm$^{-3}$}}
\def\cmmd{\hbox{\kern 0.20em cm$^{-2}$}}
\def\pc{\hbox{\kern 0.20em pc$^{2}$}}

\def\twcotwo{\hbox{${}^{12}$CO(2-1)}}

\def\thcoone{\hbox{${}^{13}$CO(1-0)}}
\def\ceio{\hbox{C${}^{18}$O}}
\def\ceiotwo{\hbox{C${}^{18}$O(2-1)}}
\def\ceioone{\hbox{C${}^{18}$O(1-0)}}

\def\csthree{\hbox{CS(3-2)}}

\def\htwo{\hbox{H${}_2$}}
\def\h13cop{\hbox{H$^{13}$CO$^{+}$}}

\begin{document}
\title{Triggered massive-star formation at the border of the H{\small II} 
region Sh~104}

\author{L.~Deharveng\inst{1}
          \and
        B.~Lefloch\inst{2}
          \and
        A.~Zavagno\inst{1}
          \and
        J.~Caplan\inst{1}
          \and
        A.P.~Whitworth\inst{3}   
          \and
        D.~Nadeau\inst{4}
          \and
        S.~Mart\'{\i}n\inst{5}
         }
  \authorrunning{L.~Deharveng et al.}
  \offprints{L.~Deharveng\\ \email{lise.deharveng@oamp.fr}}

\institute{
      Laboratoire d'Astrophysique de Marseille, 2 Place le Verrier, 13248 Marseille Cedex 4, France
       \and
        Laboratoire d'Astrophysique de l'Observatoire de Grenoble, 414 Rue de
        la Piscine, BP 53, 38041 Grenoble Cedex 9, France
       \and
      Department of Physics and Astronomy, University of Wales,
      Cardiff CF24 3YB, Wales, UK
       \and
    Observatoire du Mont Megantic et Département de Physique,
    Université de Montréal, C.P.6128, Succ.\ Centre-Ville, Montréal,
    QC, Canada, H3C3J7
       \and
    IRAM, Avenida Divina Pastora, 7, 18012 Granada, Spain
    }
\date{Received; accepted }

\abstract{
   We present millimetre-line imaging of the Galactic \HII\ region Sh~104.
   We show that it is surrounded by a ring of molecular gas and dust.
   Four large molecular condensations are regularly spaced around
   the ring. These condensations are themselves fragmented and contain
   several massive dense cores. A deeply embedded cluster is observed in
   the near IR towards the largest condensation. It contains at least
   one massive star ionizing an ultra-compact \HII\ region. The Sh~104
   region is a good illustration of the `collect and collapse' model 
   for star formation triggered by the expansion of an \HII\ region.

   \keywords{Stars: formation -- Stars: early-type -- ISM: \HII\ regions --
   ISM: individual: Sh2-104}
   }

 \titlerunning{Triggered massive star formation near Sh~104}
 \authorrunning{L.~Deharveng et al.}

\maketitle

\section{Introduction}


The expansion of an \HII\ region may trigger star formation in various
ways (see the review of Elmegreen 1998 and references therein). As an
\HII\ region expands with supersonic velocity, dense neutral material
accumulates between the ionization front and the shock front which
precedes it on the neutral side; this decelerating shocked layer may
become unstable and fragment on various time scales to form stars. This
is the `collect and collapse' model, first proposed by Elmegreen \& Lada
(1977), in which:
\begin{itemize}
  \item The layer may be unstable on a short time scale, of the order of
    the internal crossing time. The resulting stars are not massive, and
    should be observed, later on,  moving ahead of the swept-up layer.
  \item The layer may also be unstable, on a longer time scale, to
    gravitational collapse along its length. The fragments are massive
    enough to form massive stars or clusters (Whitworth et al.\ 1994),
    which should be observed in the direction of the parental layer.
\end{itemize}

It has been shown that, statistically, the more luminous protostellar
objects tend to form in molecular clouds adjacent to \HII\ regions
(Dobashi et al.\ 2001), hence the search we are carrying out for deeply
embedded massive stars and clusters at the peripheries of \HII\ regions.
We present here the case of the \HII\ region Sh~104 (Sharpless 1959). An
ultracompact (UC) \HII\ region, ionized by a deeply embedded cluster,
lies at its periphery. This cluster is possibly a second-generation
cluster, whose formation has been triggered by the expansion of Sh~104.
We present new molecular observations of this region, and discuss the
possible origin of the observed massive-star formation.

The molecular observations and the $JHK$  photometry of the cluster will
be fully presented and discussed in a forthcoming paper, hereafter
Paper~II, and a model for the star formation triggered by Sh~104 will 
be presented in Paper III.

\section{Presentation of the region}
Sh~104 is a 7\arcmin\ diameter, optically visible \HII\ region, whose
main exciting star is an O6V with a visual extinction of 4.7~mag
(Crampton et al.\ 1978, Lahulla 1985). The LSR velocity of the ionized
gas is $\sim$0~km~s$^{-1}$ (Georgelin et al.~1973). The distance to 
Sh~104 is $4.0 \pm 0.5$~kpc (see Paper~II).

Sh~104 is a thermal radio continuum source (Israel 1977). Fig.~1 shows
the radio map obtained by Fich (1993) at 1.46~GHz, with a resolution of
40\arcsec. Sh~104 exhibits a shell morphology, both at optical and radio
wavelengths, with the exciting star at the centre of the shell. Sh~104 
is a low density \HII\ region. For the assumed distance, the mass
of ionized hydrogen is $\sim450M_{\odot}$ (Israel 1977). A 
non-resolved thermal radio source lies at its eastern border, at $20^{\rm h}
17^{\rm m} 55\fs9$, $+36\degr 45\arcmin 39\arcsec$ (J2000), according to
the NVSS Source Catalog (Condon et al.\ 1998). This UC \HII\ region
coincides with the IRAS20160+3636 source discussed below. 

The Sh~104 region appears as a most remarkable object in band A of the
MSX Survey (Egan et al.\ 1999). This band covers the 6.8--10.8~$\mu$m
range with an angular resolution of $\sim20\arcsec$. Fig.~1 shows a
complete emission ring surrounding the ionized region. Band A contains
the 7.7$\mu$m and 8.6$\mu$m emission bands, often attributed to
polycyclic aromatic hydrocarbons (PAHs). This is most probably the
origin of the band A emission ring surrounding the Sh~104 ionized
region. Continuum emission due to small grains at high temperature may
also contribute to the band A emission.

An MSX point source lies in the direction of the dust ring. It is
resolved, with a HPBW (corrected for the instrumental PSF) of 21\arcsec
(0.41~pc). This MSX point source coincides with the IRAS point source
IRAS20160+3636 ($L_{\rm IR} \sim 3 \times 10^4 L_{\odot}$). A near-IR
cluster lies in the direction of this source. This cluster is most 
probably the exciting cluster of the UC \HII\ region, and therefore 
contains at least one massive star (see Paper II). We obtained
$JHK$ images of this cluster at the CFHT 3.6-m telescope in October
2002.

\begin{figure}
  \includegraphics[width=85mm]{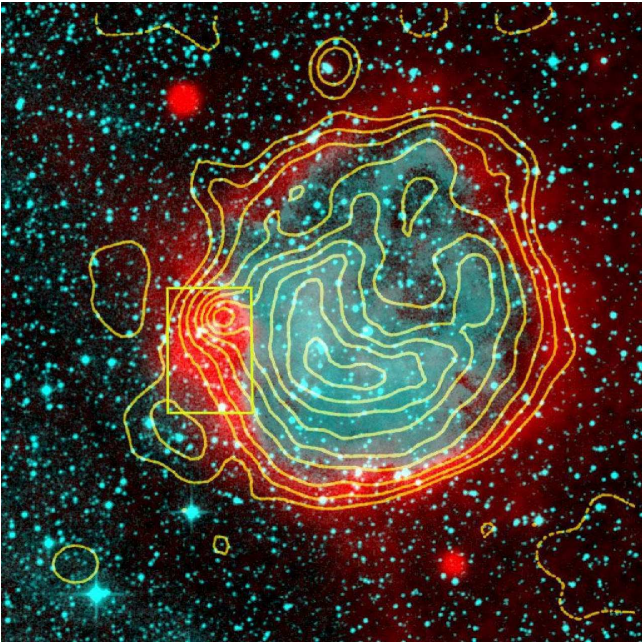}
  \caption{Composite colour image of the Sh~104 region. The emission of
  the dust appears in red (MSX survey, band A), and that of the ionized
  gas in turquoise (DSS2-red survey); the yellow contours correspond to
  the radio continuum emission at 1.46~GHz (Fich 1993). The field size
  is $12\farcm 8 \times 12\farcm 8$; north is up and east is to the
  left. The yellow rectangle shows the field of Fig.~3.}
\end{figure}

\begin{figure*}
  \includegraphics[width=180mm]{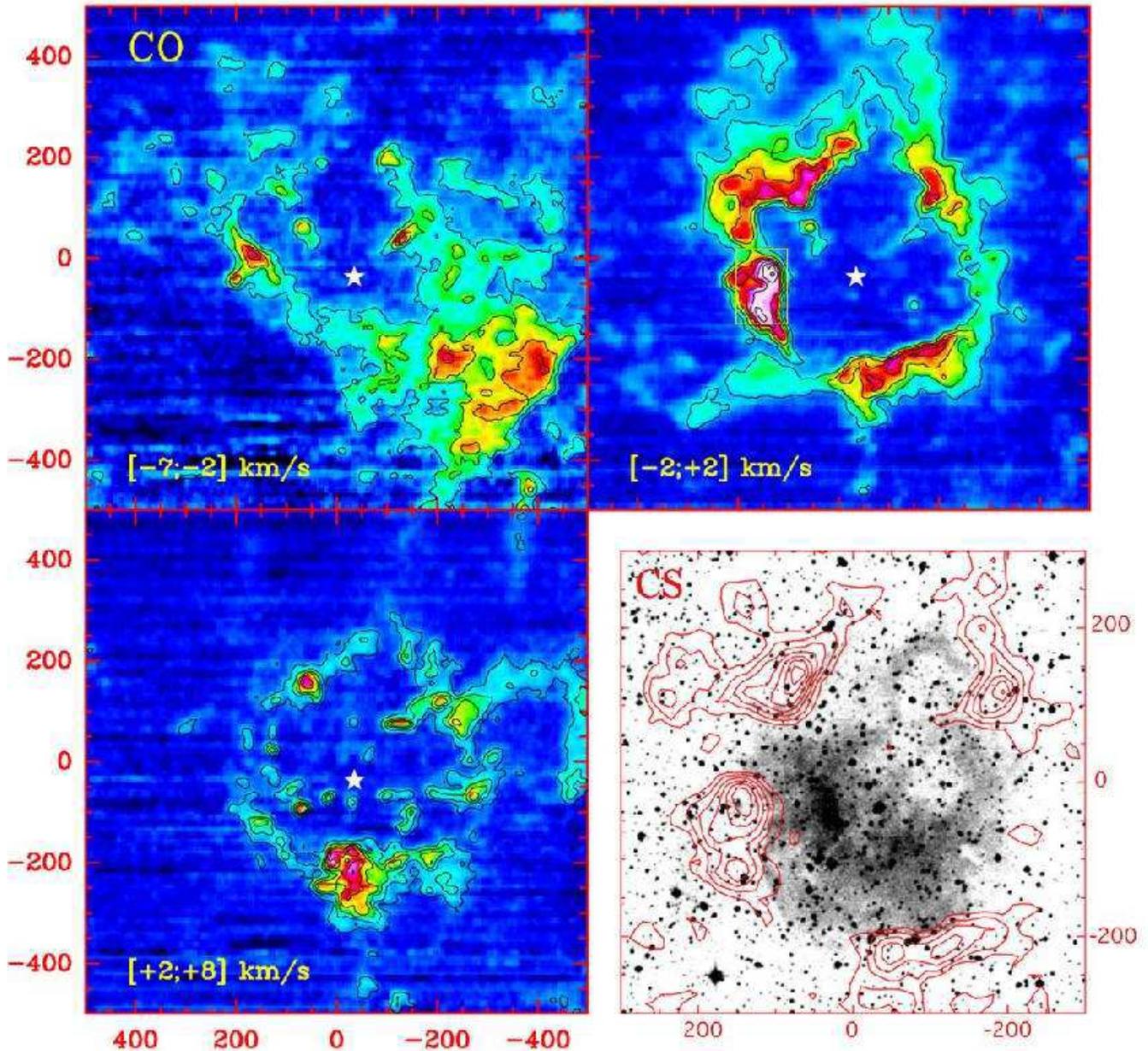}
 \caption{
{\em First three panels}: Maps of the \twcotwo\ 
velocity-integrated emission. The velocity interval is 
marked on each panel. In the $[-7;-2]~\kms$ panel the lowest contour
as well as the contour step is 5~K$\kms$; for the $[-2;+2]$ panel the 
first contour and contour interval are 10~K$\kms$; for the $[+2;+8]~\kms$ panel 
the contours levels are at 5, 10, 15, 20, 30 and 40~K$\kms$. 
Coordinates are the offsets in
arcseconds with respect to the field centre at $20^{\rm h} 17^{\rm m} 
45^{\rm s}$, $36\degr 46\arcmin 00\arcsec$ (J2000). The white 
`star' marks the
position of the exciting star. The yellow rectangle in the upper right panel 
shows the field of Fig.~3. {\em Lower right panel}: Distribution of
the CS(2-1) flux integrated between $-3$ and $+5\kms$ (red contours).
First contour and contour interval are 0.3~K$\kms$. The contours are
superposed on the DSS2-red frame of Sh~104.}
\end{figure*}


\section{Molecular observations and results}

The molecular material associated with Sh~104 was observed with the IRAM
30-m telescope in September 2002. The \twcotwo\ line emission was mapped
at 11\arcsec\ resolution with the HERA 9-channel heterodyne array in
the `on-the-fly' mode. We adopted a reference position located
$15\arcmin$ east of the exciting star. The emission at this position was
observed in frequency switching mode and subsequently added to the
observed spectra. The final map has a size of $15\arcmin \times
15\arcmin$ and is almost fully sampled. CO fluxes are given in units of
antenna temperature $T_{\rm A}^{*}$ scale). The emission of the other
tracers being limited to the denser regions, and with a much smaller
extent, we adopt the main-beam scale for brightness temperatures.

Figure~2 presents the CO emission integrated over three velocity intervals.
It shows that the CO material forms a ring which entirely
surrounds the \HII\ region. The CO brightness is highest at the border
of the ionized region, and lies in the range 17--27~K. 

The emission of the denser material was observed using the standard
heterodyne receivers in the lines of CS, HCO$^+$, $^{13}$CO and C$^{18}
$O. Fig.~2 shows that the dense gas is mainly concentrated in four
large fragments, separated from each other by 4.3 to 6.7~pc.

The brightest fragment is found in the direction of the cluster. Its
dimensions are about 3.1 pc $\times$ 1.5 pc (based on the contour at
20\% of the peak \ceioone\ intensity).  From the CO data, we
estimate a gas kinetic temperature of $\sim30$~K in the fragment. 
The mass of molecular material inside the fragment is estimated by
integrating the flux of the \ceioone\ line; assuming that the levels are
populated according to LTE and that the line is optically thin, we
derive a mass of $670 \msol$ for the fragment and a mean density of
$3100\cmmt$.

Integrating the total flux of the \thcoone\ line over the whole shell, 
and assuming the line to be optically thin with an average excitation 
temperature of 20~K, the total mass of material amounts to $6000\msol$. 

As can be seen in Fig.~2, each fragment consists of several roughly
circular subunits, or `cores', with a typical diameter of 0.4--0.6~pc
(from \csthree), marginally resolved at 2~mm (and 3~mm). Observations of
the \ceiotwo\ line show that the fragment associated with the IRAS
source is composed of three cores separated by about 0.8~pc (Fig.~3).
The stellar cluster lies between two of these cores (at about 10$\arcsec$ 
from the centre of the north-eastern core). Their physical
properties, as derived from the \ceio\ observations, are very similar.
The average gas column density is 3--3.5$\times 10^{15}\cmmd$, which
implies core masses of 70 to $100\msol$. From the \ceio\ data, we find
an average \htwo\ density of $\sim 10^4\cmmt$ inside the cores.

For the same cores, a large-velocity gradient analysis of 
three CS transitions shows that
the density of the emitting gas is $n(\htwo) =$ 1.6--2.8$\times
10^5\cmmt$ and the average column density $N(\rm CS) \simeq 10^{13}
\cmmd$. Adopting a standard abundance [CS]/[\htwo]$ = 10^{-9}$, we
derive a mass of $45 \msol$.

\begin{figure}
  \includegraphics[width=88mm]{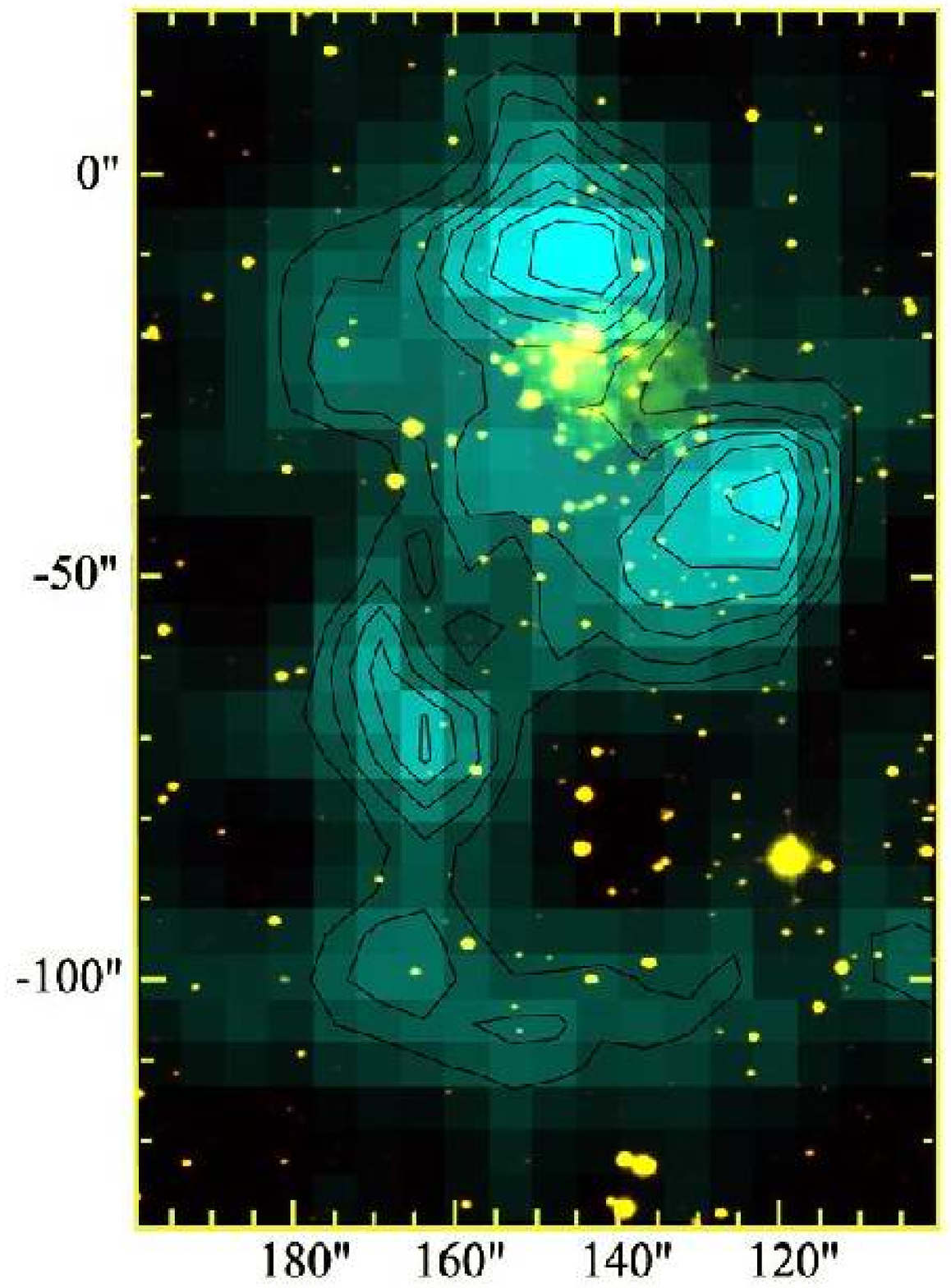}
  \caption{Map of the \ceiotwo\ emission (turquoise) of the fragment 
  associated with the cluster, integrated between radial velocities 
  $-7$ and $+8$\kms. 
  The contours are at 30\%, 40\%, 50\%, 60\%, 70\%, 
  80\% and 90\% of the peak brightness of 6.7~K$\kms$. This map is 
  superimposed on a CFHT $K$ frame of the cluster, with 
  the stars appearing in yellow. This field corresponds to the yellow 
  rectangles in Figs~1 and 2. The zero point of the coordinates is the
  same as in Fig.~2.
  }
\end{figure}

\section{Discussion and conclusions}

The existence of a molecular ring surrounding the ionized gas, and the
presence, around this ring, of {\it four} dense fragments {\it regularly}
spaced, provide strong evidence in favour of the `collect and collapse'
model. This configuration allows us to reject the following hypotheses
concerning the formation of the cluster: i)~spontaneous collapse of a
pre-existing molecular clump; ii)~collision of the compressed layer with
a pre-existing molecular clump; and iii)~collapse of a molecular
postshock core formed in the compressed layer propagating in a
supersonic turbulent medium (Elmegreen et al.\ 1995) -- in
which case the fragments would be randomly located. The cluster is most
probably a second-generation cluster, resulting from the fragmentation
of the swept-up layer due to gravitational instabilities  developing on
the long time scale, $t \sim 0.5 (G \rho _0)^{- \frac {1}{2}} \sim$ 1--3
$\times 10^{6}$~years for an \HII\ region expanding in a medium of
density $\rho _0=10^3$ to $10^2$~cm$^{-3}$, respectively. The fact that
the cluster is still observed in the direction of the compressed layer
reinforces this interpretation.

Whitworth et al. (1994) have estimated the characteristics of the 
fragments formed by gravitational instability in the dense shell 
swept up when an \HII\ region expands into an infinite, uniform, 
neutral medium. The fragment properties are insensitive to the 
rate at which ionizing photons are emitted by the central star; 
they depend primarily on the preshock density, $\rho_0$, and -- 
very strongly -- on the effective sound speed in the postshock gas, 
$a_{\rm s}$. However, with this model,  no pair of values of 
$\rho_0$ and $a_{\rm s}$ can simultaneously reproduce the 
number of fragments observed in Sh~104 (four) {\em and} their 
individual masses ($\sim 700 M_\odot$). We have therefore 
developed a new model for Sh~104.

In the new model, a massive star is formed near the centre of a 
molecular cloud of finite extent. The \HII\ region, which the massive 
star excites, expands and sweeps up the cloud into a dense shell, 
which then fragments. The model takes account of the self-gravity 
of the swept-up shell, and also the inertial drag of the material 
being swept up by the expanding shell. By invoking a cloud which 
is slightly aspherical, in the sense of being slightly more extended 
in the plane of the sky than along the line of sight, we can explain 
why all four fragments are seen -- in projection -- around the rim of 
the \HII\ region. The model will be described in detail in Paper III.

The mean density of the fragments observed around Sh~104 is high, 
$\sim$3100~cm$^{-3}$; for 
$T\sim$30~K, their Jeans length is $\sim$0.65~pc. Thus this structure
itself must fragment into several cores separated by some 0.65~pc. Such
dense cores are observed in the fragment associated with the cluster.
Due to their high density, star formation must proceed rapidly in these
dense cores ($\sim$7$\times$10$^{4}$~years for the observed density of
2$\times$10$^5$~cm$^{-3}$). This is most probably the origin of the observed
embedded cluster. The other three fragments observed at the periphery 
of Sh~104 are potential sites of massive-star formation.

The Sh~104 region appears as the prototype of massive-star formation
triggered by the expansion of an \HII\ region. Thanks to its very simple
morphology, Sh~104 is particularly helpful for understanding this
sequential star formation mechanism.

\begin{acknowledgements}
We would like to thank the IRAM \mbox{30-m} Observatory Staff for their
support during the observations, and our anonynous referee for his 
suggestions concerning the figures. This work has made use of the Simbad
astronomical database operated at CDS, Strasbourg, France. It has used
the NASA/IPAC Infrared Science Archive, which is operated by the Jet
Propulsion Laboratory, California Institute of Technology, under
contract with the National Aeronautics and Space Administration.
\end{acknowledgements}

{}

\end{document}